\documentclass[preprint,aps,showpacs]{revtex4}
\usepackage{epsfig}\usepackage{float}
\RequirePackage{amssymb,amsmath}
\usepackage{makeidx}
\usepackage[latin1]{inputenc}
\usepackage{graphicx}
\usepackage{indentfirst}
\usepackage{epsfig}
\usepackage{color}
\newcommand{\noi}{\noindent}

\begin{document}

\title{Recoil Properties of Fragments Formed in the 4.4 GeV Deuteron-Induced 
Reaction on Gold target}

\author{A. R. Balabekyan$^a$, 
N. A. Demekhina$^b$, 
G. S. Karapetyan$^c$, 
L. Karayan$^a$, 
J. R. Drnoyan$^d$, 
V. I. Zhemenik$^d$, 
J. Adam$^d$, 
L. Zavorka$^d$, A. A. Solnyshkin$^d$, V. M. Tsoupko-Sitnikov$^d$, 
J. Khushvaktov$^d$, V. Guimar\~aes$^c$}
\affiliation{
a) Yerevan State University, A. Manoogian, 1, 025, Yerevan, Armenia \\
b) Yerevan Physics Institute, Alikhanyan Brothers 2, Yerevan 0036, Armenia\\
c) Instituto de Fisica, Universidade de Sao Paulo \\
Rua do Matao, Travessa R 187, 05508-900 S~ao Paulo, SP, Brazil \\
d) Joint Institute for Nuclear Research (JINR), 
Joliot-Curie 6, Dubna 141980, Moscow region Russia\\
}

\begin{abstract}

The recoil properties of fragments produced by the interaction
of 4.4 GeV  deuteron with  $^{197}$Au target have been studied.
New experimental data on recoil properties for 90 nuclei,
varying from $^{24}$Na to $^{198}$Au, were obtained. The technique applied
was the {\it thick-target thick-catcher}  and induced activity method. 
The deuteron beam was obtained
from the Nuclotron of the Laboratory of High Energies (LHE), Joint Institute 
for Nuclear Research (JINR), Dubna. The experimental data were analyzed 
on the basis of the standard two-step vector model formalism.
From this analysis we could find evidence to support the existence of 
several different mechanisms, such as spallation, fission and 
fragmentation, in the reaction investigated. Fission contributed appreciably 
to the formation of products in the mass region of 65 $\leq$ A $\leq$ 120. 
The kinematic characteristics of residual nuclei formed in the present
deuteron-induced reaction have been compared to those 
from proton-induced reactions with gold target.
\end{abstract}


\maketitle

\section{Introduction}

In recent years, the interest to investigate kinematic characteristics 
of  reaction products has been preconditioned by the attempt to create 
an universal picture of the interaction between high energy projectiles 
with heavy nuclei targets and to determine the basic mechanism responsible for 
the development of the several process that usually occur \cite{enq,rej,ta}. 
For instance, the possibility to obtain an unified presentation of the
momentum and energy distributions of the fragments produced by  
interacting system would enrich the whole picture of reaction mechanism
and would extend the conception of the reaction models.
In relativistic nucleus-nucleus collisions, one of the most interesting 
question to be addressed is related to the energy transfer mechanism between 
the projectile and the target. Nuclear recoil experiments can provide 
valuable information, such as angular distributions and kinetic energies of
the produced nuclei. These kind of information can deepen our understanding 
of the reaction mechanism and allows the test of 
different model representations.

Recoil properties of nuclei can be  determined via the 
{\it thick-target thick-catcher} associated with the induced activity 
method.  The {\it thick-target thick-catcher} method  has been 
extensively applied to investigate hadron-induced reactions on various 
targets over a wide range of incident energy.  The important
feature in these experiments is that the thickness of the target and catcher 
foils should be larger than the longest range of the recoiled product. 
The quantities to be measured are the fractions of $F$ and $B$ 
intensities of the produced 
nuclides that recoil out of the target foil after the reaction into the 
catcher foils positioned at forward and backward directions, respectively.
Such technique has been used to investigate the fragment kinematic 
properties of proton-induced reaction in uranium target \cite{Beg,Yu1}. 
Also, for proton-nucleus on gold target, a wealth of data about the 
kinematic properties of fragments has been accumulated over the years
\cite{Kaufman1, Kaufman2, Lagarde}. These investigations were
suitable to verify the contribution of different mechanisms such as fission, 
fragmentation, and spallation in the reaction. Moreover, experiments
on kinematic properties for reaction induced by heavier projectiles,
such as high-energy $^{12}$C ions, have also been performed, where target 
fragment angular distributions were obtained \cite{Cole, Morita1, Morita2}.

In the present paper we report the results for the recoil properties 
of the fragments produced by the interaction of 4.4 GeV deuteron  with 
$^{197}$Au target. High energy reactions induced by deuteron have the 
particularity that, due to the low binding energy of  the deuteron (2.22 MeV), 
the reaction can proceed with the deuteron either as hole nucleus or as no 
interacting nucleons (proton plus neutron).  These components are revealed 
in the present experiment by considering the analysis of the kinematics 
properties of the different mass region and compare with the results from 
proton-induced reaction at similar incident energies.  We have also previously 
investigated recoil properties of fragments formed in the reactions induced 
by deuteron and proton beam at  3.65 GeV/nucleon on $^{118}$Sn 
target \cite{bal1,bal2}. We have used the same technique as in this work 
to measure the recoil properties of many of the same nuclides, thus 
permitting a comparison of how these properties vary with bombarding energy, 
target and projectile (single nucleon or system of composite nucleons).

The present paper is divided as follows; Section II is devoted to 
give some details on the experimental procedure and data analysis, 
where we have used the two-step vector model  \cite{Winsberg1, Winsberg2}.
In Section III we present the results and the discussion, where
we have performed a comparison of the results with deuteron and proton 
induced reactions. Final conclusions is given in Section IV.

\section{Experimental Procedure and data analysis}

The 4.4 GeV deuteron beam was obtained from the Nuclotron Laboratory of 
High Energies (LHE), Joint Institute for Nuclear Research (JINR), Dubna,
and used to irradiate a 39.13 mg/cm$^2$ thick gold target. The gold target
consisted of a high-purity target metal foil of  20x20 mm$^2$,
and  was sandwiched by a pair of 7.0 mg/cm$^2$ thick Mylar foil 
with the same size ({\it catcher foils}). The Mylar foils
were used  to collected the nuclei that recoiled to the forward and backward 
directions with respect to the beam. To improve the statistics  
15 gold target sandwiched by a pair of Mylar foil were piled up altogether. 
A solid angle of about  2$\pi$ is provided by using the catcher foils at the 
immediate vicinity 
of the target. The irradiation time was 28.6 
hours with a total beam intensity of about $(6.43\pm0.71)\cdot 10^{12}$ 
deuteron. The $^{27}$Al$(d,3p2n)^{24}$Na reaction with the known cross 
sections of $14.2\pm0.2$mb \cite{damdin} were used to monitor the beam.
The $\gamma$-rays from the decay of the recoiled nuclides formed in the 
target and captured in the Mylar foils (catcher foils) 
were measured, in an off-line
analysis, with High purity Germanium (HpGe) detector with 28$\%$ relative 
efficiency and an energy resolution of 2 keV ($^{60}$Co at 1332 keV). 
The measurements of $\gamma$-rays were performed over a period of time to 
follow the decay of the individual nuclide. 
The energy-dependent detection efficiency of the HpGe
detector was measured with standard calibration sources of
$^{54}$Mn, $^{57;60}$Co, $^{137}$Cs, $^{154}$Eu, $^{152}$Eu, and
$^{133}$Ba. The residual nuclides were identified by the energy,
intensity of characteristic $\gamma$-lines, and by their respective
half-lives. Nuclear properties, used to identify the observed isotopes, 
were taken from Ref. \cite{Firestone}.
The half-lives of identified isotopes were within the range of 15
min and 1 yr. In the induced activity method used here, once 
we have obtained all the $\gamma$-ray, we could determine the production 
cross sections for each fragment and reaction products.
In this particular experiment, where we used the 
{\it thick-target thick-catcher} method, we obtained the intensity 
fractions for each nuclide  recoiled 
out of a target in the forward and
backward direction,  denoted $F$ and $B$, respectively.
The uncertainties  in determining the intensity fraction for  each radionuclide
depended mainly on the following factors: the statistical significance of
the yields ($\leq$ 10\%), the accuracy in measuring the
target thickness and the accuracy of tabular data on nuclear
constants ($\leq$ 3\%), and the errors in determining the  detector
energy-dependent efficiency ($\leq$ 5\%). 

The measurements regarding the recoiling nuclei, such
as the yields in the forward and backwards directions, were transformed into
kinematic quantities by considering the two-step vector model 
 \cite{Winsberg1, Winsberg2}.
According to this method, the first stage of reaction involves the formation 
of the compound nucleus following by a cascade, leaving residual nucleus 
with an excitation energy $E^{*}$ and a velocity  $\vec{v}$ 
(or the momentum $\vec{p}$) along the beam direction. At the second stage,
the residual nucleus evaporates nucleons and/or light particles, and
as a result, the nucleus acquires an additional velocity $\vec{V}$.
Thus, the velocity $\vec{V_{l}}$ of a recoiling nuclide, in the
laboratory system, is the sum of two vectors, $\vec{V_{l}}=\vec{v}+\vec{V}$. 
The velocity vector $\vec{v}$ is the result of the fast projectile-target 
interaction, while the velocity vector $\vec{V}$, assumed to be isotropic in 
the moving system, is the result of the slow deexcitation of the excited 
primary fragment. The vector $\vec{v}$ is assumed to be constant while the 
values of the vector $\vec{V}$ are assumed to have a Maxwell distribution. 
It is also assumed that there is no correlation between the two vectors. 
Moreover, the vector $\vec{v}$ can be decomposed into its two orthogonal 
components: parallel and perpendicular to the beam, $v_{\parallel}$ and $v_{\bot}$,
respectively. Also, according to the this two-step vector model 
it was assumed that there is no correlation between the velocity 
$\vec{v}$ of the excited nucleus and $\vec{V}$, the angular distribution 
of fragments in the moving frame is isotropic, and $v_{\bot}$ is zero.

The kinematic properties of the recoiling nuclides depend on their 
range inside the target or catcher foils and their corresponding energy.  
It is convenient to express this relation as  \cite{Winsberg1, Winsberg2}:
\begin{equation}
R=kV^N ,
\label{range}
\end{equation}

\noi where $R$ is the mean range (corresponding to $V$), $k$ and $N$ are 
constants which can be evaluated from tables of ranges of nuclei recoiling 
into various materials \cite{Northcliffe}. 

As mentioned before, that main quantities  measured in this experiment are 
the fractions of  each nuclide recoiled out of the target in the forward or
backward directions. These quantities are derived as;

\begin{eqnarray}
F={S_{F}}/(S_{F}+S_{B}+S_{T}),   \qquad  B={S_{B}}/(S_{F}+S_{B}+S_{T}),
\label{fraction}
\end{eqnarray}

\noi where $S_{F}$ , $S_{B}$, and $S_{T}$ are the yields associated with the 
products formed in the target, emitted in forward or backward directions, 
and absorbed by the corresponding catcher foil.
These yields were then used to calculate the forward to backward 
($F/B$) anisotropy of the fragment emission and the ranges in the target 
material ($R$).

The ranges and the $F$ and $B$ intensities are related by the 
following expression:

\begin{eqnarray}
FW=\frac{1}{4}R [1+\frac{2}{3} (N+2) \eta + \frac{1}{4} (N+1)^2
\eta^2];  \,\,\,\,\,
 BW=\frac{1}{4} R[1-\frac{2}{3} (N+2) \eta +
\frac{1}{4} (N+1)^2 \eta^2]
\end{eqnarray}

\noindent where $\eta = v/V $ ($\eta_{\parallel}=v_{\parallel}/V$, 
is the ratio of the parallel component of the first-step to
the second-step velocity) and $W$ is the target thickness in $mg/cm^2$.

The reaction product mean ranges ($R$) were calculated using the 
follow expression  \cite{Winsberg1,Winsberg2}:

\begin{eqnarray}
 R=2W(F+B)/(1+\frac{1}{4}(N+1)^2 \eta^2)
\end{eqnarray}

With the experimental values for the $F$ and $B$ and the 
mathematical formalism developed in the two-step vector model 
\cite{Winsberg1,Winsberg2}, it was possible to  calculate parameters 
that characterize the first ($v_{\parallel}$, $E^{*}$) and the 
second ($R$ and $T$) stages of the 
interaction, where $T$ is the kinetic energy of a fragment and $E^{*}$ is the 
mean excitation energy after cascade nucleus.

The relative velocity of the formed fragment, $v_{\parallel}/v_{CN}$, where
$v_{CN}$ is the velocity of a hypothetical compound nucleus formed in a 
complete fusion, is the main feature that can be considered as the sign of a 
complete or an incomplete fusion.

The value of the forward velocity $v$ may be used to determine the
average cascade deposition energy (excitation energy, $E^*$)
as follows:
\cite{schd}:

\begin{eqnarray}
E^{*}=3.253*10^{-2}k^{'}A_t v [T_p/(T_p+2)]^{0.5} ,
\end{eqnarray}

\noindent where $E^*$  and the bombarding energy $T_p$ are expressed in
terms of $m_pc^2$. $A_t$ is the target mass in $amu$ and $v$ is in
units of (MeV/$amu)^{0.5}$. The constant $k'$ has been evaluated
by  Scheidemann and Porile \cite{schd} on the basis of Monte Carlo 
cascade calculations as $k^{'}=0.8$.

\section{Results and Discussion}

The experimental values of $F/B$ ratio are given in Table I for each
nuclide observed in the present experiment.
The $F/B$ values represent somehow the extent of the recoiling nuclei 
at the forward direction, and thus can be considered an indirect measurement 
of the forward momentum  transferred to the target nucleus in the reaction. 
The variation of $F/B$ as a function of the product mass $A$ is shown in 
Fig. 1. As it can be seen in this figure, $F/B$ shows a peak at 
the high mass  region which decreases as the mass loss from the compound 
nucleus increases, until about 20 nucleons have been lost. With further mass 
loss it goes to the deep spallation region (lower mass region).  
There is a broad minimum in the mass region of $ A = 45-75$, and then $F/B$  
increases again as one goes to the light fragment region.  A similar mass 
dependence has been observed previously for the 1-11.5 GeV proton bombardment 
on Au target  \cite{Kaufman1}. This mass dependence could be explained by 
invoking different mechanisms for the production of nuclei in different mass 
regions.  For high-energy induced reaction, as the case of the present 
experiment, heavy residual nuclei are produced mainly via spallation 
mechanism, with fragments preferably leaving at forward direction.
On the other hand, the isotropic distribution of light nuclei may be due to  
fragmentation or fission-like processes.

The recoil parameter $2W(F+B)$ is related to
the  mean ranges of the recoiling nuclei in the target material.
Actually, the mean range of
the recoils is somewhat smaller than $2W(F+B)$, but it is convention
to refer to the latter as the range. The obtained values of this parameter 
for each nuclide studied in this experiment are listed in Table I and
plotted as a function of the mass number $A$
of the fragments in Fig. 2. A smooth curve has been drawn 
over the data just to indicate the trend.
The nuclides with higher mass, close
to the mass of the compound system, is expected to be produced by the 
spallation process; and since it is a peripheral interaction, the values 
for the mean range, $2W(F+B)$, should be small, as observed.
Intermediate mass range nuclides were formed mainly by deep spallation 
and fission-like mechanisms, and have a relatively larger range associated 
with a larger contribution of their binary decay. The particles with 
highest ranges are the light fragments $^{24}$Na and $^{28}$Mg. 
Such fragments can be produced, for instance, in (multi)fragmentation process 
with low impact parameter in the collision.

By analyzing the contents of Table I and II,
some discussion about the different processes that can occur in the 
reaction of  4.4 GeV deuteron with gold can be made.
The heaviest fragments ($A \geq 131$) show large mean values of
$F/B\sim 3$ and very low fragment kinetic energies ($\sim$ 0.03
MeV/nucleon). These products are the result of spallation of the
$^{197}$Au target. The products in the  intermediate mass range of 
$65 \leq A \leq 120$  have the average value of $F/B = 1.64$ and 
mean kinetic energy of $\sim$ 0.22 MeV/nucleon. The lower $F/B$ values obtained 
for these products in this mass range can be an evidence  for fission process
contribution.  However, we can not clearly isolated the fission process 
contribution from the contribution of  other processes (as deep spallation) 
in this mass range region.
The lightest fragments, $^{24}$Na and $^{28}$Mg ($A < 40$), are characterized 
by high kinetic energies and relatively large values of $F/B$ and, thus, 
their production is consistent with a "multifragmentation" mechanism
\cite{Wolfgang}. 
The intermediate mass fragments ($40 < A < 65$) represent 
the group of fragments with relatively high kinetic energies ($\sim$ 0.56
MeV/nucleon) and  large values of $F/B = 1.50$.
Such fragments could be associate with fission where 
the heavy partner fragments would be  in the mass  region 
of $A \sim 120-130$. 
Deep spallation process may also contribute with products in this
mass range. Deep spallation process can be responsible not only for
the emission of nucleons and light charge particles (with $Z\leq 2$) but 
also for emission of these relatively light fragments, specially 
if we take into account their  high kinetic energies and formation in 
non peripheral collision  (high values of $F/B$).

The interesting question concerning the mechanism involved 
in relativistic nucleus-nucleus collision refers on how the 
energy transfer proceeds, whether the kinetic
energy is transferred from the projectile to the target as a whole system
or as energy per nucleon. To answer to this  
we compared our results of a deuteron induced
reaction with those from reaction induced by high energy protons. 
In Figs. 3 and 4, we show such comparison for
recoil properties of fragments for the reaction induced
by  4.4 GeV deuteron and similar
measurements for the same products from the reaction induced by
1.0, 3.0 and 11.5 GeV protons with also $^{197}$Au target \cite{Kaufman1}.

In Fig. 3  we present the comparion of the forward-to-backward ratio,
$F/B$, for the deuteron 
and proton induced reactions at different energies. 
As mentioned before, nuclides at intermediate to high mass
range are formed by spallation process which corresponds to 
a more peripheral collision. These heavier nuclides are formed with higher 
probability at lower projectile incident energies. 
Thus, by increasing the projectile energy, the spallation reaction probability 
decreases \cite{kaufman3}. As can be observed in Fig. 3 (a), 
the ratio of  $F/B$ for 4.4 GeV deuteron and 1.0 GeV proton for 
the mass range of nuclides formed by deep spallation process are similar 
except for the spallation products (mass range of $140<A<180$), where
the $(F/B)_{deuteron}/(F/B)_{proton}$ ratio drops off. The dropping of this ratio
in this mass region is not observed in Figs. 3 (b) and (c) where
the proton bombarding energy is higher, 3.0 GeV and 11.5 GeV, 
respectively. Actually, as the proton energy is further increased to 11.5 GeV, 
the  $(F/B)_{deuteron}/(F/B)_{proton}$ ratio are systematically larger. 
This can be explained by the domination of a non peripheral collisions 
with a smaller impact parameter for high-energy projectile. 
By observing the trend of the data in Fig. 3 (b), we can say
that target fragment kinematic properties from deuteron-induced
reaction mostly resemble those from the reaction induced by protons
at the same total projectile energy. This observation is 
confirmed by theoretical calculations by Cugnon in Ref. \cite{Cugnon} and it is
consistent with the data of Kaufman \cite{Kaufman1}, which
found that the recoil properties of target fragment from reaction induced 
by  25.2 GeV $^{12}$C beam on $^{197}$Au are similar to those  
from the reaction induced by 28 GeV protons on $^{197}$Au target.

The other recoil property that can be used to compare the present
data with those induced by protons at different incident energy
is the recoil parameter, $2W(F+B)$, which is 
related to the mean range of the nuclides. Such comparison
is presented in Figs. 4 (a), (b) and (c). As can be observed, the
ratio of the values of $2W(F+B)$ for products from both deuteron
and proton induced reactions (especially the one at 3.0 GeV proton),
as a function of the mass number are similar. This indicates that the
deexcitation phase of these reactions is similar for this energy
range.

A discussion on the transferred momentum and kinetic energy of
the fragments can also be made. The two-step vector model applied 
here assumes that all the forward momentum transferred from
the incident particle occurs in the first step of the reaction
($v_{\parallel}$), and all of the isotropic processes occur in the 
second step ($V$, $T$), where the kinetic energy $T$ 
is derived from range-energy data \cite{Northcliffe}.
We listed in Table II the values of the kinetic energy ($T$), 
the parameter $N$, as well as $P (=AV)$  values for each nuclide,
where $P$ is the mean momentum imparted to the target fragment with
mass number $A$ during the deexcitation step of the reaction. 
The mean momentum $P$ as a function of the mass number
of the residual nuclides is present in Fig. 5. The dependence of $P$ 
on the $\Delta{A^{0.5}}$ (the square root of the number of nucleons 
removed from the target) is also indicated in this figure as the solid line 
curve. As one may observe (within certain broad
limits) there is a general dependence of $P$ upon $\Delta{A^{0.5}}$. The same
tendency was found for the 8 GeV $^{20}$Ne with $^{181}$Ta system 
\cite{Loveland}, which produces the same compound nucleus
as the present work, and in the proton-induced reactions for 
the deep spallation product range $A=140-200$ \cite{Crespo}. 
According to the basic assumptions of the two-step vector model 
for high energy reactions, this dependence is an indication of a 
sequential, stepwise momentum kicks being imparted to the fragment 
during the deexcitation phase of the  reaction. Also, the similarity of the 
$P$ values for the fragments produced by  different projectiles
 asserts  that the deexcitation phase of these reactions is the same and 
does not depend on the type of projectile.

Due to correlation between $v_{\parallel}$ and $V$  we can estimate 
the parallel velocity component transferred to an intermediate nucleus in 
the first cascade step ($v_{\parallel}$) and from the  eq. (5) we can estimate 
the mean excitation energy of  the residual nucleus after cascade nucleus 
($E^{*}$). The longitudinal velocity 
$v_{\parallel}$ for the recoiling nuclei are presented in Table II
and plotted in Fig. 6.  The presence of a plateau at wide range of 
fragments can be explained as possible saturation of the energy-momentum 
transfer in a nuclear collision. 
For comparison, we also plotted in Fig. 6 the longitudinal velocity 
$v_{\parallel}$  of the products from reactions induced by protons 
 with 1.0, 3.0 and 11.5 GeV \cite{Kaufman1}. As we can 
observe in the figure, there is a gradual decrease
of longitudinal velocity with the increase of the total projectile
energy.  Our data follow well the general tendency of
$v_{\parallel}$ values for protons. This behaviour suggests that
a similar process takes place after the first step for reaction 
induced by different kind of incident 
particles at similar incident energy. This also may indicate 
a similar process for transfering the forward momentum.
In summary, we can say that in the reactions induced by projectiles, with
equivalent incident energy, on heavy targets, the first stage
of the reaction and the deexcitation of the fragments following the
initial fast stage, are carried out in a similar manner.


The variation of excitation energy ($E^{*}$) after cascade
nucleus as a function of the mass number of fragments is shown in Fig. 7. 
For the medium-mass nuclides, the excitation energy is higher for 
the lighter nuclides, for example for $^{24}$Na and $^{28}$Mg nuclides
(two outstanding points in the figure). Also, these nuclides requires 
the largest excitation energy for their formation  mechanism. 
The residual nuclides in the mass range of $65 \leq A \leq 120$ have
about the same excitation energy, indicating that
fission and probably deep spallation process take place at
the same excitation energy regime. The products of spallation, the ones
farther from the target mass, require more energy to be produced. 
The excitation energy for spallation products would be smaller 
if they were formed only by evaporation of light particles up to $^{4}$He.
We divided the total mass range in regions and estimated, on the basis of 
eq. (5), the corresponding mean excitation energy. The obtained values 
consisted of 1.416$\pm$283 MeV (for the region of $A < 40$); 
472$\pm$95 MeV (for $42 \leq A \leq 59$); 
353$\pm$71 MeV (for $65 \leq A \leq 120$); 264$\pm$53 MeV (for $A \geq 131$).
This large variation of values can be understood
by considering that different mechanism are involved in the 
formation of these residual nuclides \cite{Aleklett,sugar,Yu2}.

\section{Conclusion}

Kinematic properties of reaction products from the interaction
of  4.4 GeV deuteron on  $^{197}$Au target were investigated  for the first
time. The experimental results were analyzed  on the basis of the two-step 
vector model  and discussed in terms of different reaction mechanisms, 
such as spallation, fission, deep spallation and fragmentation. 

The dependence of the recoil properties 
on the mass number of the product has been studied. 
The variation of the forward to backward intensity ratio, $F/B$, 
and mean recoil range, $2W(F+B)$, with the mass of the products
show that heavier products were formed probably in a more
peripheral collisions with large impact parameter, with 
relative low excitation energy, while medium and light mass products 
might have been produced in a more central collisions with smaller 
impact parameter and high excitation energy of the reaction remnant. 
The  linear momentum transferred to an intermediate nucleus ($p_{\parallel}$)  
shows a gradual decrease with the increase of the projectile energy. 
The combination of the information on the deposited energy, 
$E^*$, and on the forward cascade momentum allow us to suggest the 
existence of different mechanism in the formation of the products in
this deuteron-induced reaction.

The dependence of recoil properties 
on mass number of the product has been also compared to those from 
reactions induced by energetic protons on the same gold target.
The similarity of the general picture of the mass dependence of 
the kinematic parameters concerning the first step (intranuclear cascade) and
second (evaporation ) stage is a clear indication that
the interaction of protons and deuteron at the same  energies
are very similar. This behavior can be explained by the absence of 
additional interaction between the two nucleons of the deuteron with the 
target nucleus. This fact has been confirmed  also by our
previous work concerning the study of the cross sections of the
deuteron and proton induced reactions on Sn isotopes \cite{bal2}.


\section{Acknowledgment} G. Karapetyan is grateful to Funda\c c\~ao
de Amparo \`a Pesquisa do Estado de S\~ao Paulo (FAPESP)
2011/00314-0 and 2013/01754-9, and also to
International Centre for Theoretical Physics (ICTP) under the
Associate Grant Scheme. The authors are grateful to group leader 
of LNP JINR Dr. S. Avdeev for granting the possibility of carrying out 
this experiment, and also to the lead researcher of LHEP JINR Kh. 
Abraamyan for the help during the experiment.

\newpage
\begin{table}
\caption{Target fragment recoil properties. $W$ is the thickness of gold foil
and thus $2W(F+B)$ has unit of $mg/cm^2$.}
\vskip 0.5cm
\begin{tabular}{|c|c|c|c|c|c|c|c|c|} \hline
Nuclide & $F/B$ & $2W(F+B)$ & Nuclide & $F/B$ & $2W(F+B)$ & Nuclide & $F/B$ & $2W(F+B)$\\
\hline
$^{24}$Na  & 1.94$\pm$0.29 & 15.7$\pm$2.3 & $^{86}$Rb  & 1.67$\pm$0.38 & 5.1$\pm$1.1   & $^{135}$I  & 2.69$\pm$0.61 & 1.52$\pm$0.35\\
\hline
$^{28}$Mg  & 1.94$\pm$0.29 & 14.5$\pm$2.1 & $^{86}$Y   & 1.68$\pm$0.25 & 5.09$\pm$0.76 & $^{135}$Ce & 2.90$\pm$0.43 & 1.45$\pm$0.21\\
\hline
$^{42}$K   & 1.50$\pm$0.56 &  8.0$\pm$1.3 & $^{87}$Y   & 1.56$\pm$0.23 & 5.59$\pm$0.83 & $^{139}$Ce & 3.44$\pm$0.51 & 1.41$\pm$0.21\\
\hline
$^{43}$K   & 1.57$\pm$0.23 &  7.9$\pm$1.1 & $^{88}$Y   & 1.67$\pm$0.25 & 4.37$\pm$0.65 & $^{143}$Pm & 2.94$\pm$0.49 & 1.39$\pm$0.23\\
\hline
$^{44m}$Sc & 1.65$\pm$0.25 &  7.2$\pm$1.0 & $^{89}$Zr  & 1.61$\pm$0.24 & 5.01$\pm$0.75 & $^{145}$Eu & 3.80$\pm$0.57 & 1.26$\pm$0.18\\
\hline
$^{46}$Sc  & 1.54$\pm$0.23 &  8.3$\pm$1.2 & $^{90}$Nb  & 1.72$\pm$0.25 & 4.57$\pm$0.68 & $^{146}$Eu & 3.78$\pm$0.86 & 1.37$\pm$0.31\\
\hline
$^{47}$Ca  & 1.47$\pm$0.22 &  8.2$\pm$1.2 & $^{93}$Mo  & 1.65$\pm$0.24 & 3.97$\pm$0.59 & $^{146}$Gd & 3.21$\pm$0.73 & 1.46$\pm$0.33\\
\hline
$^{48}$Sc  & 1.66$\pm$0.25 &  8.2$\pm$1.2 & $^{93}$Tc  & 1.60$\pm$0.24 & 4.02$\pm$0.60 & $^{147}$Eu & 2.94$\pm$0.44 & 1.43$\pm$0.21\\
\hline
$^{48}$V   & 1.55$\pm$0.23 &  7.0$\pm$1.0 & $^{95}$Zr  & 1.66$\pm$0.38 & 4.03$\pm$0.92 & $^{147}$Gd & 3.25$\pm$0.48 & 1.38$\pm$0.20\\
\hline
$^{51}$Cr  & 1.44$\pm$0.22 &  7.4$\pm$1.1 & $^{95}$Nb  & 1.74$\pm$0.40 & 4.8$\pm$1.1   & $^{148}$Eu & 2.90$\pm$0.43 & 1.33$\pm$0.19\\
\hline
$^{52}$Mn  & 1.65$\pm$0.28 &  5.9$\pm$1.0 & $^{96}$Tc  & 1.68$\pm$0.25 & 4.66$\pm$0.69 & $^{149}$Eu & 3.16$\pm$0.47 & 1.15$\pm$0.17\\
\hline
$^{54}$Mn  & 1.28$\pm$0.19 &  7.5$\pm$1.1 & $^{97}$Ru  & 1.96$\pm$0,29 & 4.50$\pm$0.67 & $^{149}$Gd & 3.64$\pm$0.54 & 1.19$\pm$0.17\\
\hline
$^{55}$Co  & 1.30$\pm$0.19 &  6.7$\pm$1.0 & $^{99}$Mo  & 1.73$\pm$0.25 & 4.01$\pm$0.60 & $^{151}$Tb & 3.74$\pm$0.56 & 1.20$\pm$0.18\\
\hline
$^{56}$Co  & 1.35$\pm$0.31 &  7.1$\pm$1.6 & $^{103}$Ru  & 1.50$\pm$0.22 & 5.59$\pm$0.83 & $^{155}$Dy & 3.80$\pm$0.57 & 1.16$\pm$0.17\\
\hline
$^{58}$Co  & 1.53$\pm$0.22 &  6.3$\pm$0.9 & $^{104}$Ag  & 1.58$\pm$0.23 & 5.22$\pm$0.78 & $^{157}$Dy & 3.67$\pm$0.55 & 1.13$\pm$0.16\\
\hline
$^{59}$Fe  & 1.40$\pm$0.32 &  7.3$\pm$1.6 & $^{105}$Ag  & 1.70$\pm$0.25 & 5.21$\pm$0.78 & $^{167}$Tm & 3.51$\pm$0.59 & 0.75$\pm$0.12\\
\hline
$^{65}$Zn  & 1.34$\pm$0.22 &  5.4$\pm$0.9 & $^{110}$In  & 2.13$\pm$0.31 & 3.18$\pm$0.47 & $^{171}$Lu & 3.09$\pm$0.46 & 0.60$\pm$0.09\\
\hline
$^{71}$As  & 1.52$\pm$0.22 &  8.7$\pm$1.3 & $^{111}$Ag  & 1.62$\pm$0.24 & 4.88$\pm$0.73 & $^{173}$Hf & 3.56$\pm$0.53 & 0.67$\pm$0.10\\
\hline
$^{72}$Zn  & 1.44$\pm$0.21 &  6.7$\pm$1.0 & $^{111}$In  & 2.23$\pm$0.33 & 3.79$\pm$0.56 & $^{175}$Hf & 2.88$\pm$0.48 & 0.61$\pm$0.10\\
\hline
$^{72}$As  & 1.51$\pm$0.22 &  7.6$\pm$1.1 & $^{112}$Pd  & 1.70$\pm$0.25 & 4.79$\pm$0.71 & $^{182}$Os & 3.10$\pm$0.46 & 0.44$\pm$0.06\\
\hline
$^{73}$Se  & 1.52$\pm$0.22 &  7.1$\pm$1.0 & $^{113}$Sn  & 1.60$\pm$0.24 & 4.30$\pm$0.64 & $^{183}$Re & 2.59$\pm$0.44 & 0.51$\pm$0.08\\
\hline
$^{74}$As  & 1.38$\pm$0.20 &  7.3$\pm$1.1 & $^{115}$Ag  & 1.70$\pm$0.39 & 4.14$\pm$0.95 & $^{186}$Ir & 2.70$\pm$0.40 & 0.23$\pm$0.03\\
\hline
$^{75}$Se  & 1.54$\pm$0.23 &  5.0$\pm$0.7 & $^{115m}$Cd & 1.59$\pm$0.23 & 4.62$\pm$0.69 & $^{188}$Ir & 2.50$\pm$0.57 & 0.22$\pm$0.05\\
\hline
$^{76}$Br  & 1.49$\pm$0.22 &  5.6$\pm$0.8 & $^{117m}$Sn & 1.62$\pm$0.24 & 3.77$\pm$0.56 & $^{188}$Pt & 2.80$\pm$0.47 & 0.23$\pm$0.03\\
\hline
$^{77}$Br  & 1.44$\pm$0,21 &  6.5$\pm$0.9 & $^{120m}$Sb & 1.84$\pm$0.42 & 3.00$\pm$0.69 & $^{190}$Ir & 2.60$\pm$0.59 & 0.20$\pm$0.04\\
\hline
$^{81}$Rb  & 1.54$\pm$0.23 &  6.4$\pm$0.9 & $^{131}$Ba & 2.90$\pm$0.43 & 2.90$\pm$0.43  & $^{191}$Pt & 2.30$\pm$0.34 & 0.19$\pm$0.02\\
\hline
$^{82}$Br  & 1.55$\pm$0.23 &  6.6$\pm$1.0 & $^{132}$Te & 3.033$\pm$0.45 & 2.00$\pm$0.30 & $^{194}$Au & 2.10$\pm$0.31 & 0.12$\pm$0.02\\
\hline
$^{83}$Rb  & 1.54$\pm$0.23 &  6.2$\pm$0.9 & $^{132}$Ce & 2.75$\pm$0.41 & 1.66$\pm$0.24  & $^{196}$Au & 2.00$\pm$0.30 & 0.07$\pm$0.01\\
\hline
$^{84}$Rb  & 1.61$\pm$0.24 &  6.0$\pm$0.9 & $^{133}$I & 2.67$\pm$0.40 & 1.55$\pm$0.23   & $^{197m}$Hg & 2.25$\pm$0.51 & 0.11$\pm$0.03\\
\hline
$^{85}$Sr  & 1.84$\pm$0.27 &  5.6$\pm$0.8 & $^{134}$I & 2.43$\pm$0.36 & 1.60$\pm$0.24   & $^{198}$Au & 2.23$\pm$0.37 & 0.097$\pm$0.016\\
\hline
\end{tabular}
\end{table}

\begin{table}
\caption{Target fragment kinematic properties as deduced from the 
two-step vector model. The kinetic energy $T$ in in unit of MeV, the 
momentum $P$ is in unit of  $(MeV\cdot a.m.u.)^{1/2}$ and 
$v_{\parallel}$ is in unit of $(MeV/a.m.u)^{1/2}$ }
\begin{tabular}{|c|c|c|c|c|c|c|c|c|c|} \hline
Nuclide&$N$&$T$&$P[=AV]$&$v_{\parallel}$&Nuclide&$N$&$T$&$P[=AV]$&$v_{\parallel}$\\
 \hline
$^{24}$Na  & 1.59 & 47.4$\pm$7.1 & 47.5$\pm$7.1 & 0.21$\pm$0.03 & $^{86}$Rb & 1.10 & 17$\pm$2 & 54$\pm$12 & 0.08$\pm$0.01 \\
\hline
$^{28}$Mg  & 1.47 & 45.0$\pm$6.8 & 51.1$\pm$7.7 &  0.29$\pm$0.04 & $^{86}$Y &  1.20 &  18.7$\pm$2.8 & 56.1$\pm$8.1   & 0.08$\pm$0.01\\
\hline
$^{42}$K   & 1.15 & 27.0$\pm$4.0 & 47.6$\pm$8.0 &  0.12$\pm$0.02 & $^{87}$Y &  1.20 &  21.7$\pm$3.2 & 60.9$\pm$9.1   & 0.08$\pm$0.01\\
\hline
$^{43}$K   & 1.13 & 26.0$\pm$3.9 & 47.3$\pm$7.1 &  0.12$\pm$0.02 & $^{88}$Y &  1.20 &  14.2$\pm$2.1 & 49.3$\pm$7.3   & 0.07$\pm$0.01\\
\hline
$^{44m}$Sc & 1.09 & 24.9$\pm$3.7 & 46.1$\pm$6.9 &  0.13$\pm$0.02 & $^{89}$Zr &  1.13 & 18.1$\pm$2.7 & 56.7$\pm$8.5   & 0.08$\pm$0.01\\
\hline
$^{46}$Sc  & 1.12 & 30.9$\pm$4.6 & 53.4$\pm$8.0 &  0.12$\pm$0.02 & $^{90}$Nb &  1.15 & 15.9$\pm$2.4 & 53.7$\pm$8.0   & 0.08$\pm$0.01\\
\hline
$^{47}$Ca  & 1.09 & 28.5$\pm$4.3 & 59.9$\pm$8.8 &  0.10$\pm$0.01 & $^{93}$Mo &  1.28 & 13.2$\pm$1.9 & 48.3$\pm$7.2   & 0.07$\pm$0.01\\
\hline
$^{48}$Sc  & 1.12 & 33.4$\pm$5.0 & 52.9$\pm$7.9 &  0.15$\pm$0.02 & $^{93}$Tc &  1.18 & 13.6$\pm$2.0 & 50.6$\pm$7.6   & 0.06$\pm$0.01\\
\hline
$^{48}$V   & 1.10 & 25.2$\pm$3.8 & 49.2$\pm$7.3 &  0.11$\pm$0.02 & $^{95}$Zr &  1.13 & 11.7$\pm$1.7 & 47.6$\pm$10.9  & 0.06$\pm$0.01\\
\hline
$^{51}$Cr  & 1.09 & 28.5$\pm$4.3 & 59.1$\pm$8.8 &  0.10$\pm$0.01 & $^{95}$Nb &  1.15 & 17.0$\pm$2.5 & 55.6$\pm$12.8  & 0.08$\pm$0.01\\
\hline
$^{52}$Mn  & 1.08 & 19.5$\pm$2.9 & 45.2$\pm$7.6 &  0.11$\pm$0.02 & $^{96}$Tc &  1.18 & 17.1$\pm$2.5 & 57.0$\pm$8.5   & 0.08$\pm$0.01\\
\hline
$^{54}$Mn  & 1.09 & 29.5$\pm$4.4 & 56.6$\pm$8.8 &  0.07$\pm$0.01 & $^{97}$Ru &  1.14 & 16.2$\pm$2.4 & 56.6$\pm$8.4   & 0.10$\pm$0.01\\
\hline
$^{55}$Co  & 1.08 & 26.5$\pm$4.0 & 54.1$\pm$8.1 &  0.07$\pm$0.01 & $^{99}$Mo &  1.28 & 13.0$\pm$1.9 & 49.6$\pm$7.4   & 0.07$\pm$0.01\\
\hline
$^{56}$Co  & 1.08 & 29.0$\pm$4.3 & 57.0$\pm$13.2&  0.08$\pm$0.01 & $^{103}$Ru & 1.14 & 22.6$\pm$3.4 & 67.4$\pm$10.1  & 0.07$\pm$0.01\\
\hline
$^{58}$Co  & 1.07 & 22.3$\pm$3.3 & 50.9$\pm$7.6 &  0.09$\pm$0.01 & $^{104}$Ag & 1.24 & 22.2$\pm$3.3 & 67.1$\pm$10.0  & 0.07$\pm$0.01\\
\hline
$^{59}$Fe  & 1.07 & 27.4$\pm$4.1 & 57.1$\pm$13.1 & 0.08$\pm$0.01 & $^{105}$Ag & 1.24 & 22.0$\pm$3.3 & 67.0$\pm$10.0  & 0.08$\pm$0.01\\
\hline
$^{65}$Zn  & 1.11 & 17.7$\pm$2.6 & 43.7$\pm$7.4 &  0.05$\pm$0.01 & $^{110}$In & 1.21 & 10.1$\pm$1.5 & 49.1$\pm$7.3   & 0.08$\pm$0.01\\
\hline
$^{71}$As  & 1.06 & 45.4$\pm$6.8 & 75.5$\pm$11.3 & 0.11$\pm$0.02 & $^{111}$Ag & 1.24 & 19.1$\pm$2.8 & 63.8$\pm$9.5   & 0.07$\pm$0.01\\
\hline
$^{72}$Zn  & 1.08 & 23.7$\pm$3.5 & 58.2$\pm$7.7 &  0.07$\pm$0.01 & $^{111}$In & 1.21 & 13.4$\pm$2.0 & 55.7$\pm$8.3   & 0.10$\pm$0.01\\
\hline
$^{72}$As  & 1.06 & 36.4$\pm$5.4 & 67.6$\pm$10.1 & 0.10$\pm$0.02 & $^{112}$Pd & 1.24 & 17.9$\pm$2.6 & 61.7$\pm$9.2   & 0.07$\pm$0.01\\
\hline   
$^{73}$Se  & 1.08 & 32.1$\pm$4.8 & 68.3$\pm$10.2 & 0.10$\pm$0.01 & $^{113}$Sn & 1.28 & 17.2$\pm$2.5 & 62.0$\pm$9.3   & 0.06$\pm$0.01\\
\hline
$^{74}$As  & 1.08 & 31.9$\pm$4.7 & 65.3$\pm$9.8 &  0.08$\pm$0.01 & $^{115}$Ag & 1.24 & 14.3$\pm$2.1 & 56.6$\pm$13.0  & 0.06$\pm$0.01\\
\hline
$^{75}$Se  & 1.12 & 31.3$\pm$4.6 & 49.6$\pm$7.4 &  0.10$\pm$0.01 & $^{115m}$Cd & 1.25 & 17.8$\pm$2.6 & 63.9$\pm$9.5  & 0.06$\pm$0.01\\
\hline
$^{76}$Br  & 1.08 & 21.1$\pm$3.1 & 55.8$\pm$8.3 &  0.07$\pm$0.01 & $^{117m}$Sn & 1.28 & 13.8$\pm$2.0 & 56.6$\pm$8.4  & 0.06$\pm$0.01\\
\hline
$^{77}$Br  & 1.08 & 26.6$\pm$3.9 & 63.8$\pm$9.5 &  0.08$\pm$0.01 & $^{120m}$Sb & 1.31 &  9.8$\pm$1.4 & 46.1$\pm$10.6 & 0.06$\pm$0.01\\
\hline
$^{81}$Rb  & 1.10 & 26.9$\pm$4.0 & 65.2$\pm$9.7 &  0.09$\pm$0.01 & $^{131}$Ba  & 1.26 &  9.8$\pm$1.4 & 54.2$\pm$8.1  & 0.10$\pm$0.01\\
\hline 
$^{82}$Br  & 1.08 & 26.4$\pm$3.9 & 65.6$\pm$9.8 &  0.09$\pm$0.01 & $^{132}$Te  & 1.27 &  5.0$\pm$0.8 & 42.3$\pm$6.3  & 0.07$\pm$0.01\\
\hline
$^{83}$Rb  & 1.10 & 26.1$\pm$3.9 & 65.0$\pm$9.7 &  0.08$\pm$0.01 & $^{132}$Ce  & 1.87 &  6.3$\pm$01.0 & 40.9$\pm$6.1 & 0.08$\pm$0.01\\
\hline
$^{84}$Rb  & 1.10 & 23.7$\pm$3.5 & 62.9$\pm$9.4 &  0.09$\pm$0.01 & $^{133}$I   & 1.32 &  3.6$\pm$0.6  & 37.2$\pm$5.5 & 0.06$\pm$0.01\\
\hline
$^{85}$Sr  & 1.13 & 19.6$\pm$2.9 & 57.7$\pm$8.6 &  0.10$\pm$0.01 & $^{134}$I   & 1.32 &  3.8$\pm$0.6  & 38.3$\pm$5.5 & 0.05$\pm$0.01\\
\hline
\end{tabular}
\end{table}

\newpage
\begin{table}
\caption{Continuation of Table II.}
\begin{tabular}{|c|c|c|c|c|c|c|c|c|c|} \hline
Nuclide&$N$&$T$&$P[=AV]$&$v_{\parallel}$&Nuclide&$N$&$T$&$P[=AV]$&$v_{\parallel}$\\
\hline
$^{135}$I  & 1.32 & 3.54$\pm$0.53 & 37.2$\pm$8.5 & 0.05$\pm$0.01 & $^{167}$Tm & 2.02 & 3.51$\pm$0.59 & 34.2$\pm$5.8 & 0.05$\pm$0.01\\
\hline
$^{135}$Ce & 1.87 & 5.51$\pm$0.83 & 38.5$\pm$5.7 & 0.07$\pm$0.01 & $^{171}$Lu & 2.07 & 2.97$\pm$0.44 & 31.8$\pm$4.7 & 0.04$\pm$0.01\\
\hline
$^{139}$Ce & 1.87 & 5.30$\pm$0.79 & 38.3$\pm$5.7 & 0.08$\pm$0.01 & $^{173}$Hf & 2.05 & 3.36$\pm$0.50 & 31.7$\pm$4.7 & 0.06$\pm$0.01\\
\hline
$^{143}$Pm & 1.90 & 5.53$\pm$0.83 & 39.7$\pm$6.7 & 0.07$\pm$0.01 & $^{175}$Hf & 2.05 & 3.05$\pm$0.46 & 30.5$\pm$5.2 & 0.05$\pm$0.01\\
\hline
$^{145}$Eu & 1.96 & 5.25$\pm$0.79 & 38.9$\pm$5.8 & 0.09$\pm$0.01 & $^{182}$Os & 2.14 & 2.42$\pm$0.36 & 29.9$\pm$4.4 & 0.04$\pm$0.01\\
\hline
$^{146}$Eu & 1.96 & 5.72$\pm$0.86 & 42.0$\pm$9.6 & 0.09$\pm$0.01 & $^{183}$Re & 2.19 & 2.76$\pm$0.41 & 32.2$\pm$5.4 & 0.04$\pm$0.01\\
\hline
$^{146}$Gd & 1.90 & 6.25$\pm$0.94 & 42.6$\pm$9.8 & 0.08$\pm$0.01 & $^{186}$Ir & 2.13 & 1.35$\pm$0.20 & 22.7$\pm$3.4 & 0.030$\pm$0.004\\
\hline
$^{147}$Eu & 1.97 & 5.97$\pm$0.90 & 42.4$\pm$6.3 & 0.07$\pm$0.01 & $^{188}$Ir & 2.13 & 1.32$\pm$0.20 & 22.5$\pm$5.1 & 0.030$\pm$0.004\\
\hline
$^{147}$Gd & 1.90 & 5.89$\pm$0.88 & 41.5$\pm$6.2 & 0.08$\pm$0.01 & $^{188}$Pt & 2.14 & 1.38$\pm$0.21 & 23.1$\pm$3.9 & 0.030$\pm$0.004\\
\hline
$^{148}$Eu & 1.96 & 5.55$\pm$0.83 & 40.4$\pm$6.0 & 0.07$\pm$0.01 & $^{190}$Ir & 2.13 & 1.18$\pm$0.18 & 21.5$\pm$4.9 & 0.030$\pm$0.004\\
\hline
$^{149}$Eu & 1.96 & 4.78$\pm$0.72 & 37.9$\pm$5.6 & 0.07$\pm$0.01 & $^{191}$Pt & 2.14 & 1.15$\pm$0.17 & 27.3$\pm$4.1 & 0.020$\pm$0.003\\
\hline
$^{149}$Gd & 1.90 & 5.03$\pm$0.75 & 38.1$\pm$5.7 & 0.08$\pm$0.01 & $^{194}$Au & 2.15 & 0.81$\pm$0.11 & 18.1$\pm$2.7 & 0.020$\pm$0.002\\
\hline
$^{151}$Tb & 1.94 & 5.19$\pm$0.78 & 39.0$\pm$5.8 & 0.08$\pm$0.01 & $^{196}$Au & 2.15 & 0.50$\pm$0.07 & 14.0$\pm$2.1 & 0.010$\pm$0.002\\
\hline
$^{155}$Dy & 1.94 & 5.18$\pm$0.78 & 40.1$\pm$6.0 & 0.08$\pm$0.01 & $^{197m}$Hg & 2.22 & 0.76$\pm$0.17 & 17.3$\pm$3.8 & 0.020$\pm$0.003\\
\hline
$^{157}$Dy & 1.95 & 4.99$\pm$0.75 & 39.6$\pm$5.9 & 0.08$\pm$0.01 & $^{198}$Au & 2.15 & 0.63$\pm$0.09  & 15.6$\pm$2.6 & 0.020$\pm$0.002\\
\hline
\end{tabular}
\end{table}

\newpage

\begin{figure}
\epsfig{file=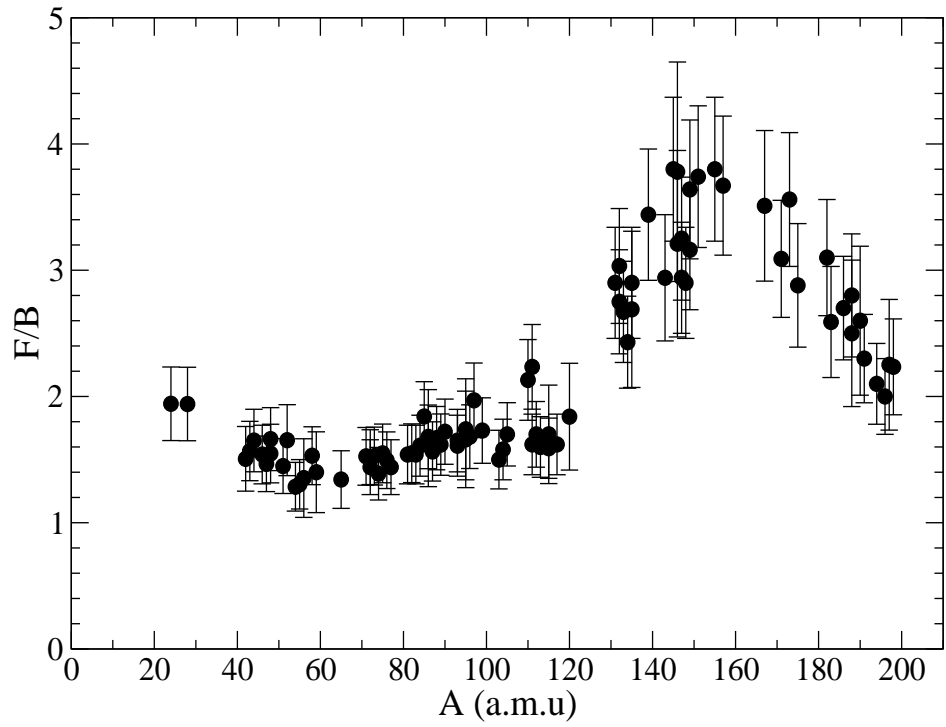,height=15cm,width=12cm,angle=-90.}
\caption{Ratio of forward to backward intensities, $F/B$, as
a function of the product mass number.}
\end{figure}

\begin{figure}
\epsfig{file=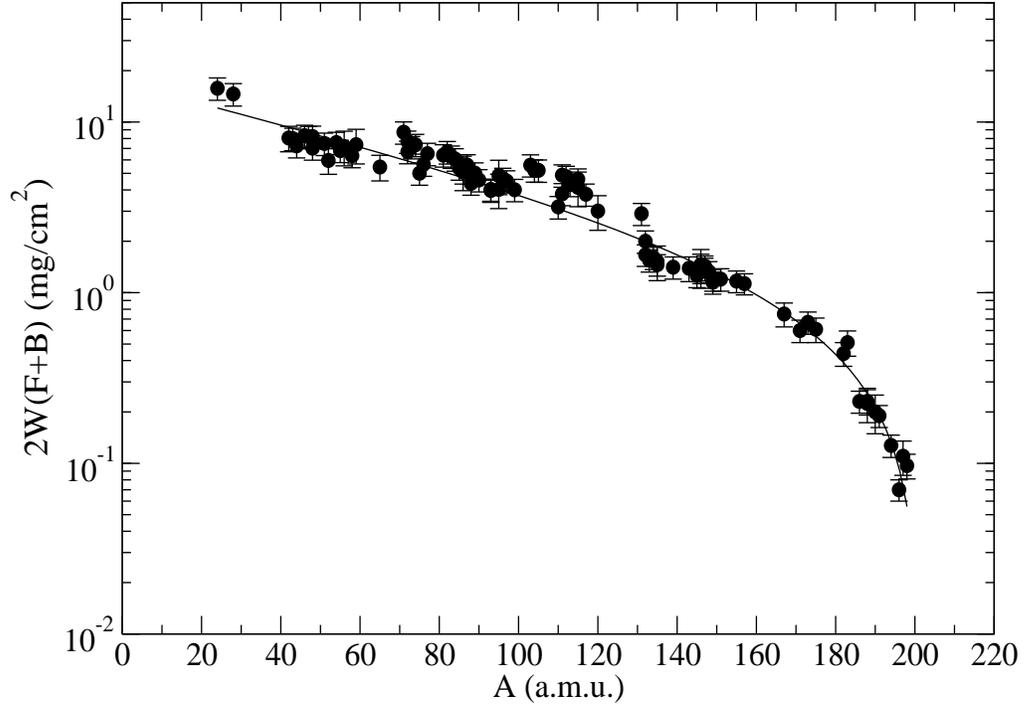,height=15cm,width=12cm,angle=-90.}
\caption{The mean range, $2W(F+B)$, as a function of the 
product mass number. The curve shows the general trend of the data.}
\end{figure}

\begin{figure}
\epsfig{file=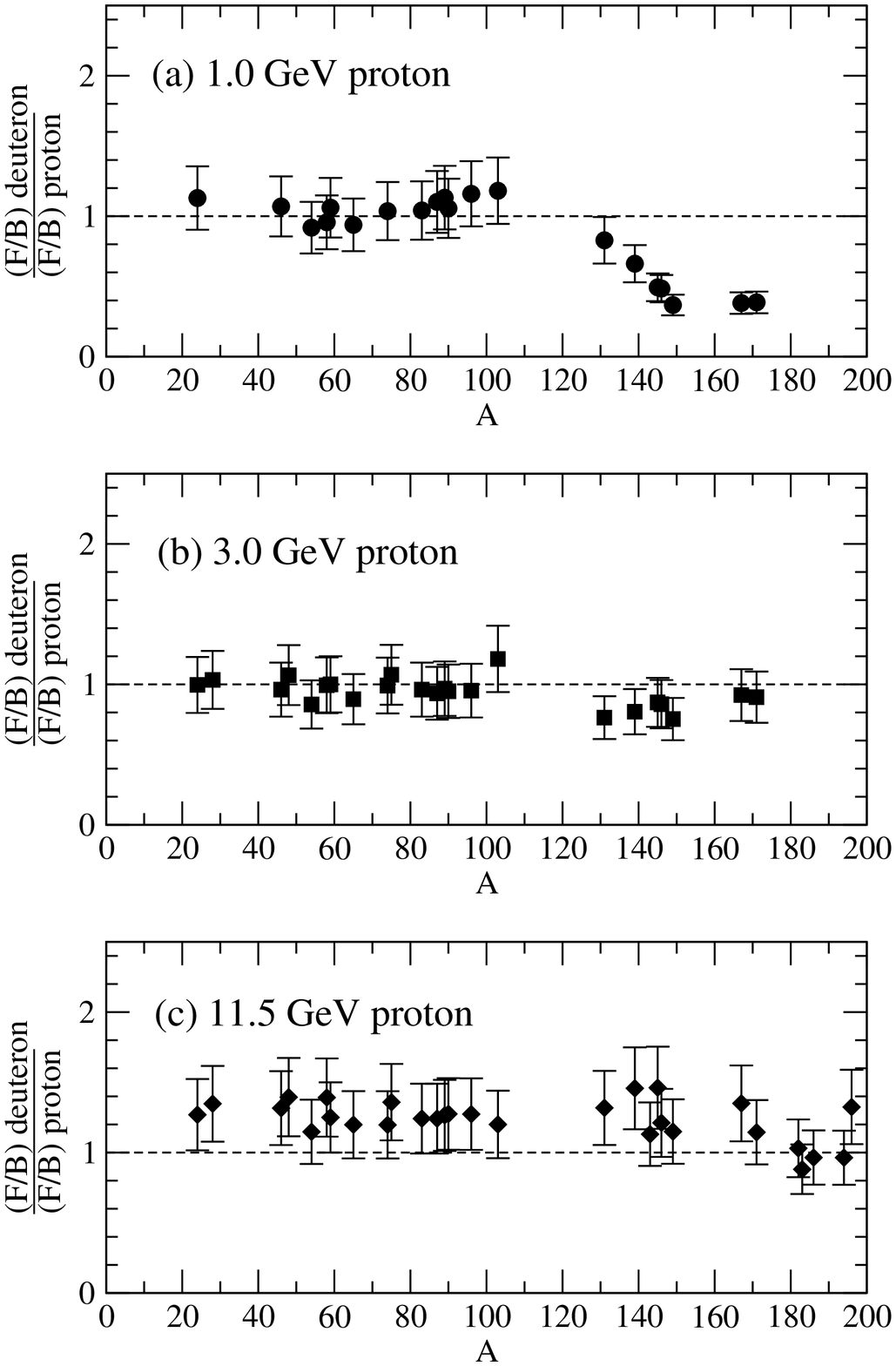,height=15cm,width=12cm,angle=0.}
\caption{The ratio of $F/B$ values from the present work and 
those measured for the reaction induced by 
1.0 GeV, 3.0 GeV and 11.5 GeV  protons  from Ref. \cite{Kaufman1}, as indicated.}
\end{figure}

\begin{figure}
\epsfig{file=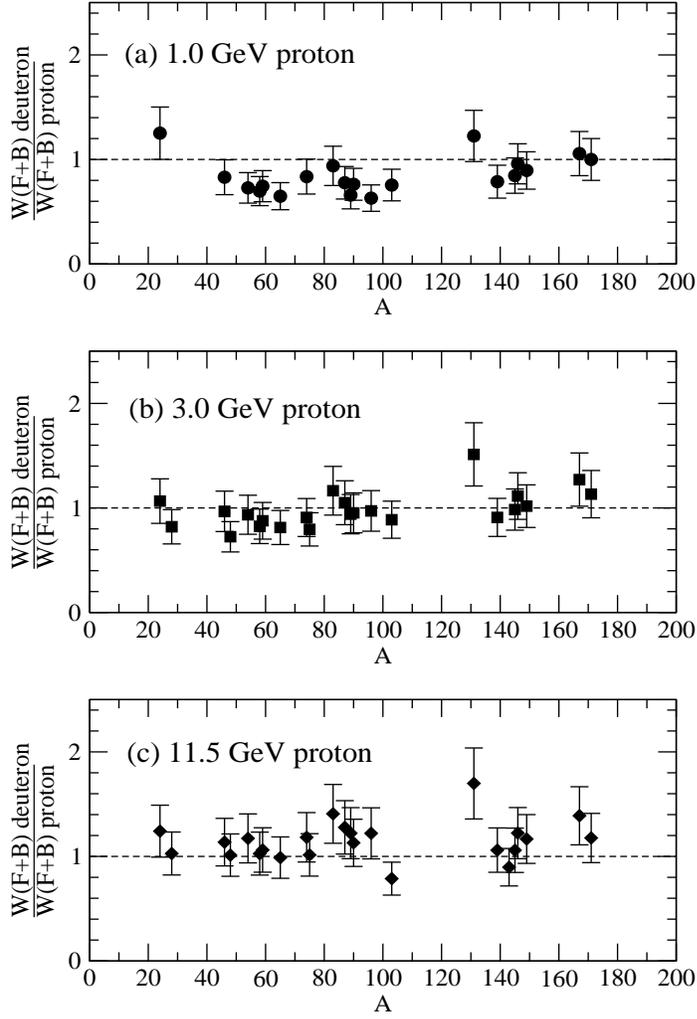,height=15cm,width=12cm,angle=0.}
\caption{The ratio of the $2W(F+B)$ values from the present work 
and those measured for the reaction induced by  
1.0 GeV, 3.0 GeV and 11.5 GeV  protons from Ref.  \cite{Kaufman1}, as indicated,
as a function of the target fragment mass.}
\end{figure}

\begin{figure}
\epsfig{file=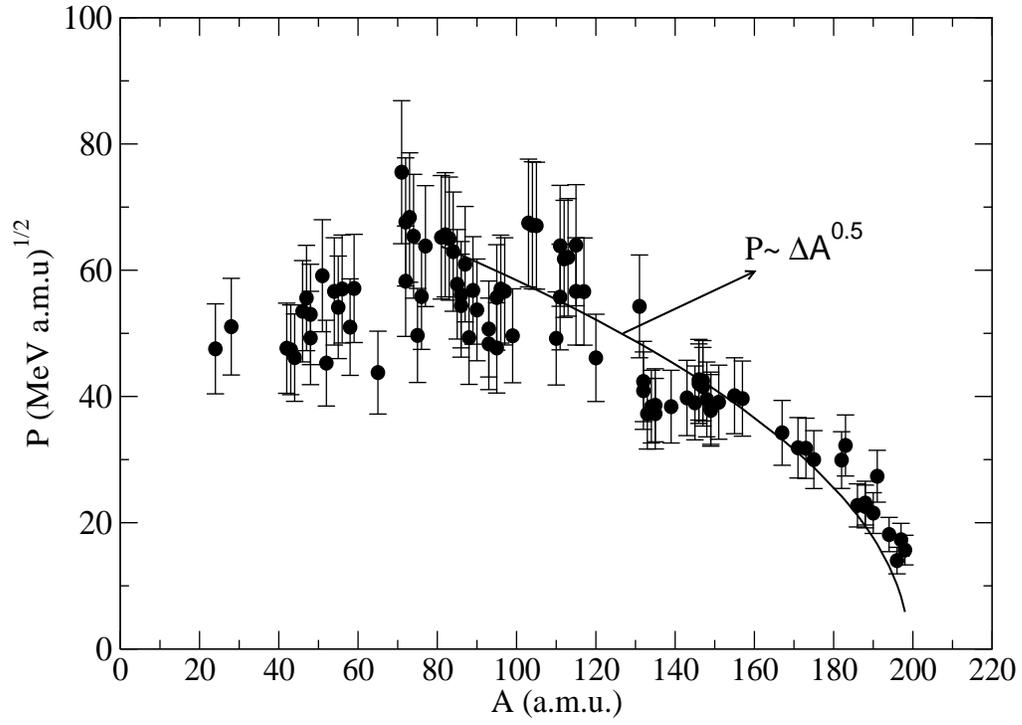,height=15cm,width=12cm,angle=-90.}
\caption{The mean momentum, $P=AV$, as a function of 
product mass number. Solid line is $P \sim \Delta A^{0.5}$ dependence.}
\end{figure}

\begin{figure}
\epsfig{file=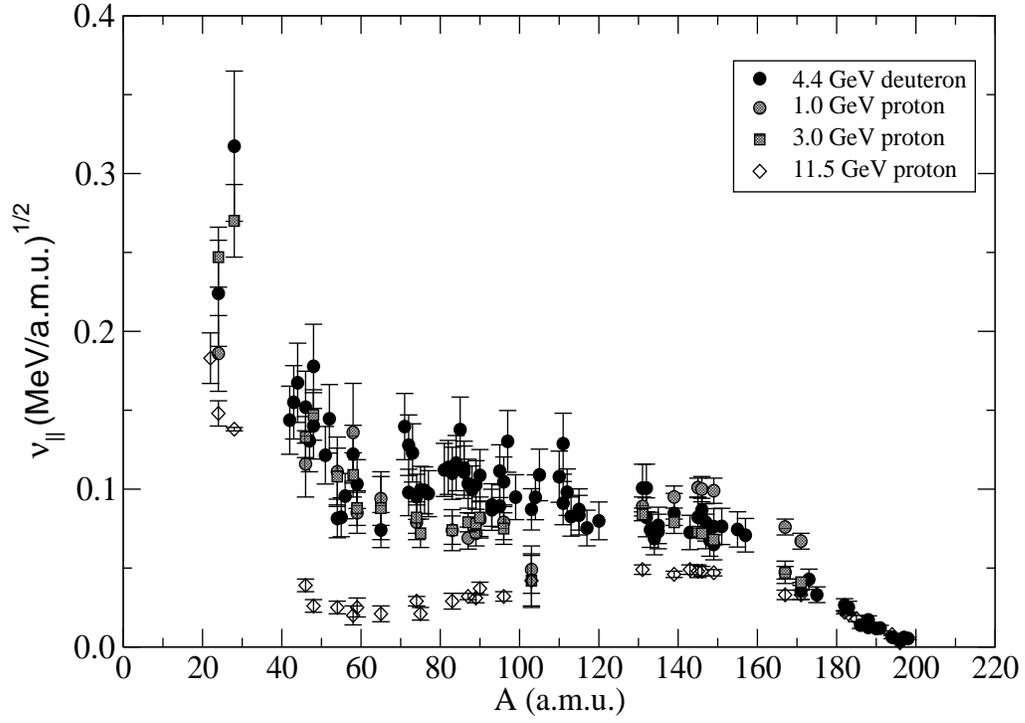,height=15cm,width=12cm,angle=-90.}
\caption{The longitudinal velocity of residual
nuclides after cascade nuclei,  $v_{\parallel}$, as a function
of ther mass number, $A$. The black circle are data from 
the present work.  Gray and open symbols are data from proton-induced 
reaction on Au \cite{Kaufman1}, with energies as indicated.}
\end{figure}

\begin{figure}
\epsfig{file=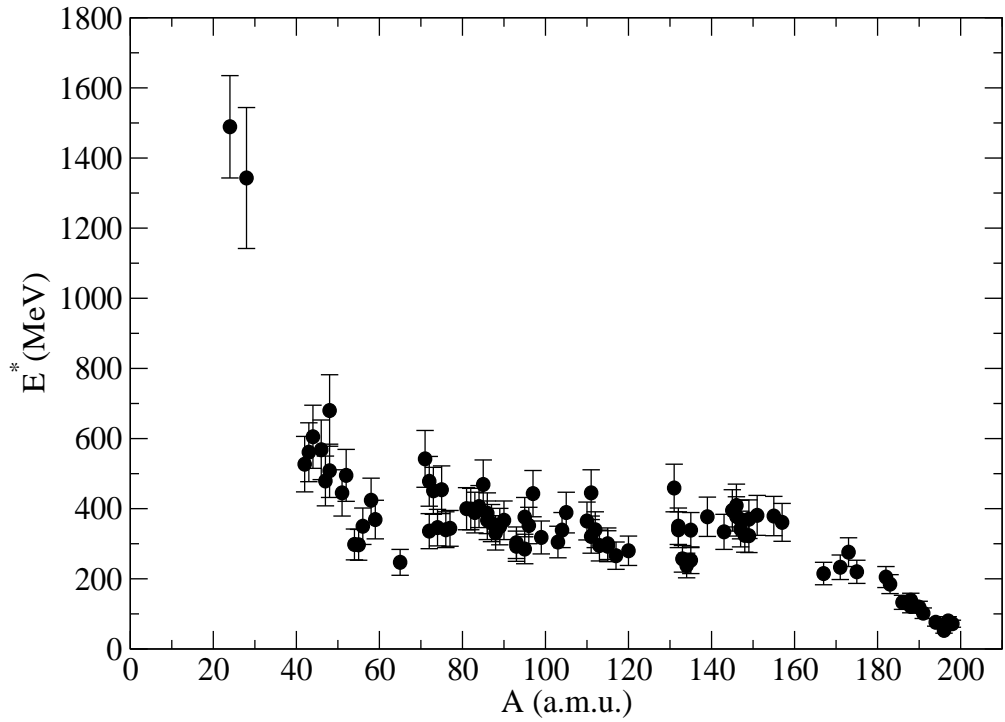,height=15cm,width=12cm,angle=-90.}
\caption{Mean excitation energy of the
residual  cascade nucleus ($E^{*}$) as a function of the product mass number.}
\end{figure}

\end{document}